\documentclass{moriond}

\def\Journal#1#2#3#4{{#1} {\bf #2}, #3 (#4)}

\newcommand{\Photo}{\includegraphics[height=35mm]{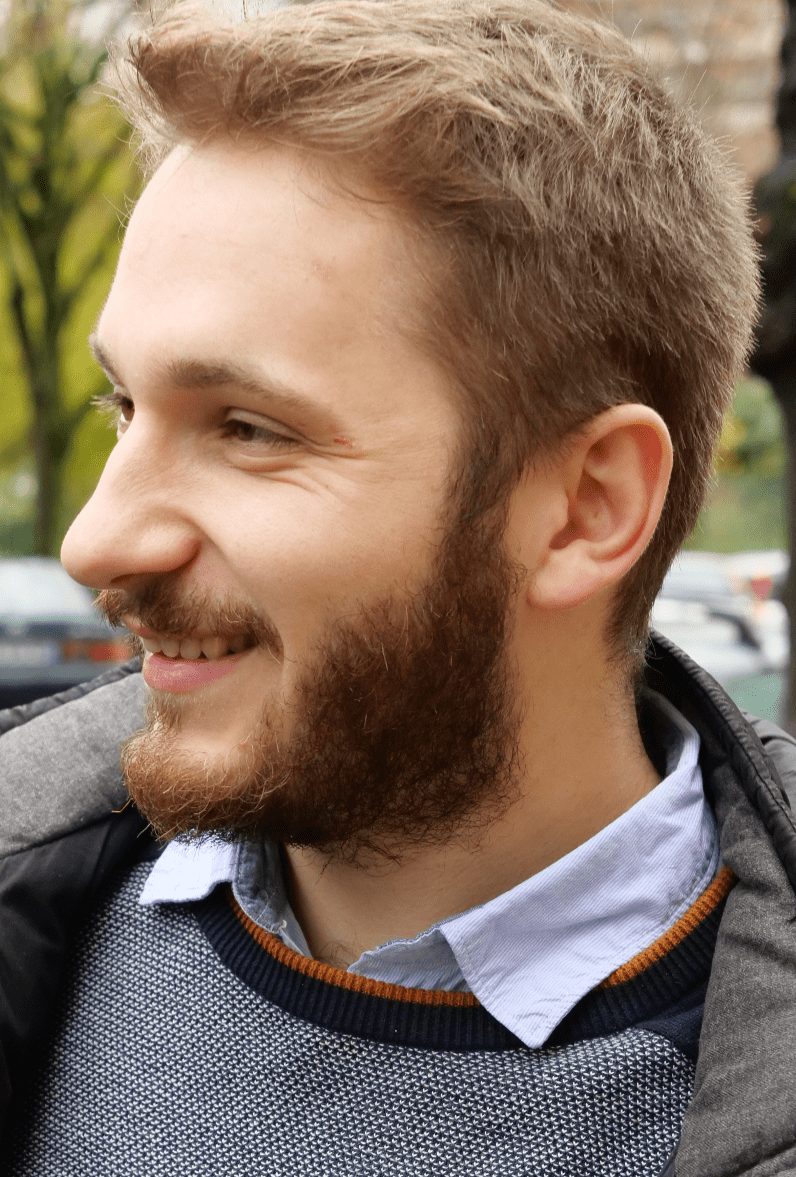}}

\begin{document}
\vspace*{4cm}
\title{Lyman-alpha forest tomography and cross-correlation with cosmic voids}

\author{ Corentin Ravoux }

\address{IRFU, CEA, Universit{\'e} Paris-Saclay, F-91191 Gif-sur-Yvette, France}

\maketitle\abstracts{
The Lyman-alpha (Ly$\alpha$) forest is a unique probe of large-scale matter density fluctuations at high redshift $z>2$. It is possible to obtain 3D maps of the matter distribution from Ly$\alpha$~data, using tomographic reconstruction methods. Here, we present the largest tomographic map of matter fluctuations at $z>2$, over the Gpc$^3$ volume covered by Ly$\alpha$~forest from SDSS-IV quasar spectra in the Stripe 82 field~\cite{tomo}. We present a catalog of high-redshift voids constructed from this map. The measurement of the cross-correlation between these voids and the Ly$\alpha$~forest provides the first observation of the matter velocity flow around voids, through the RSD effect, at such high redshift. The data is in good agreement with simulations and is well adjusted with a  linear, Kaiser velocity model~\cite{crosscorr}}

\section{Lyman-\texorpdfstring{$\alpha$}{α} tomography}

The Ly$\alpha$~forest is a tracer of neutral hydrogen in the cosmic web. It is most easily observed in quasar spectra. When observed from ground-based telescopes at a redshift $z>2$, quasar spectra show a broad peak of Ly$\alpha$ emission at $\lambda_{\rm rest} = 1215$\AA. Bluewards of this peak, a forest of lines corresponds to the absorption of light by the intergalactic medium (IGM) located between the quasar and the observer. These absorption features constitute the Ly$\alpha$~forest and on large scale they trace the neutral hydrogen in the IGM. The measurement of the Ly$\alpha$~forest is complicated when using noisy spectra. As a first step, the product of the continuous emission of the quasar $C_{q}$ by the average fraction of transmitted flux $\overline{F}$ is measured. From a quasar flux $f(\lambda)$, it is then possible to define the Ly$\alpha$~absorption contrast~\cite{bao}:

\begin{equation}
    \label{eq:delta_flux_continuum}
    \delta_{F}(\lambda) = \frac{f(\lambda)}{C_{q}(\lambda_{\mathrm{rf}})\overline{F}(\lambda)} - 1
\end{equation}

Standard Ly$\alpha$~BAO analysis, e.g.\cite{bao}, use $\delta_{F}$ to calculate correlations. In this study, we interpolate between different quasar lines-of-sight to create a 3D map of Ly$\alpha$~absorption. This is called Ly$\alpha$~tomography~\cite{pichon}. It was achieved for the first time from observations by the CLAMATO collaboration~\cite{clamato} on a portion of the COSMOS field. This measurement was on a small, dense field of quasars and Lyman-break galaxies. The ultimate goal in that case, by improving the map resolution down to a $\sim 1~\mathrm{Mpc}.h^{-1}$ scale, is to trace the filaments of the cosmic web. This is expected to be achievable with future telescopes of the ELT class.

\begin{figure}
\begin{minipage}{1.0\linewidth}
\centerline{\includegraphics[trim=2cm 1.5cm 1.5cm 3.5cm, width=0.85\linewidth]{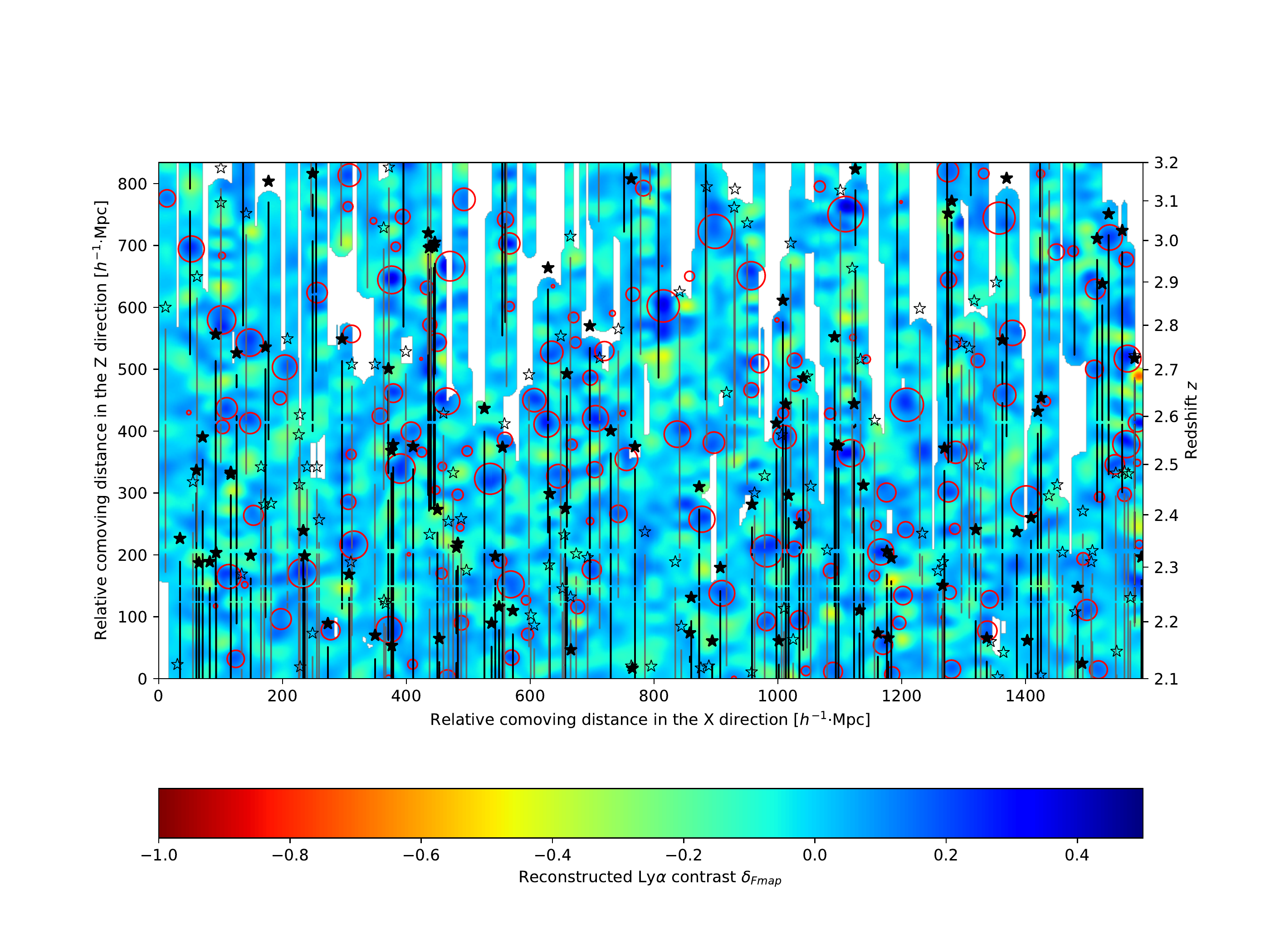}}
\end{minipage}
\caption[]{Slice of the tomographic map computed from the observed eBOSS Ly$\alpha$~forest in the Stripe 82. The slice is at constant declination $\delta_{J2000} = 0^{\circ}$, and covers right ascension $1\leq \alpha \leq 23^{\circ}$, roughly a quarter of the Stripe 82 field. The reconstruction length is $13~\mathrm{Mpc}.h^{-1}$. Circles are the intersection between the represented slice and identified voids. Filled (empty) stars represent quasars whose distance along the $y$ axis is less than 5 (10) $\mathrm{Mpc}.h^{-1}$ from the slice. Lines-of-sight used for the tomographic reconstruction are pictured as full (dotted) lines if they are 5 (10) $\mathrm{Mpc}.h^{-1}$ from the slice.}
\label{fig:tomo}
\end{figure}

Our study focused on building a tomographic map on a much larger volume, at the price of degraded resolution. We used the Ly$\alpha$~forest region from quasar spectra available in the 16$^{th}$ Data Release of the SDSS-IV eBOSS survey. In particular, we focused on the densest and most homogeneous part of this survey: a narrow band called Stripe 82, of 220 deg$^2$ area, located in the equatorial plane. We used the spectra from 8200 quasars, corresponding to a surface density of 37 quasars per deg$^2$. To build a tomographic map, we applied a Wiener filter as implemented by CLAMATO~\cite{clamato}. This algorithm performs an interpolation with a Gaussian kernel taking into account the Ly$\alpha$~forest noise in individual pixels. We included a series of useful tools for large-volume tomography in a python package \texttt{lelantos}~\cite{lelantos}. Fig.~\ref{fig:tomo} represents a slice of the tomographic map computed from eBOSS data. The main parameter of the tomographic algorithm is the correlation length of the Gaussian kernel. We chose it to be $13~\mathrm{Mpc}.h^{-1}$, which corresponds to the average separation between lines-of-sight. This tomographic reconstruction constitutes the first large-volume, high-redshift 3D map of matter density fluctuations.

In parallel, we used synthetic data called mocks to test the tomographic reconstruction algorithms. These log-normal mocks are computed using the fluctuating Gunn-Peterson approximation~\cite{weinberg}. They also provide the underlying matter density field associated with the Ly$\alpha$~absorption contrast: we could therefore use them to quantify by how much the Ly$\alpha$ tomographic map traces the matter density. We find that the correlation between these two fields is 34\%.

The tomographic map obtained from the eBOSS data can be used for several applications. First, a stack of the tomographic map around quasars reveals a clear signal centered on the quasar position: this is a recast view of the cross-correlation between quasars and the Ly$\alpha$~forest as studied in~\cite{bao}. Then, we identified eight proto-cluster candidates by selecting overdensities in the map, and requiring they are crossed by a large number of lines-of-sight. Finally, the application we focused most is the search for voids. Indeed, since our tomographic map traces matter fluctuations on large scales $\geq 13\mathrm{Mpc}.h^{-1}$, it is well adapted to search for voids, the largest structures in the cosmic web. With a spherical void finder, that we have developed in the \texttt{lelantos}~\cite{lelantos} package, we created the largest catalog of large voids at high redshift. To build this catalog, we selected only voids crossed by several lines-of-sight, and whose radius is larger than $7~\mathrm{Mpc}.h^{-1}$.

\section{Ly\texorpdfstring{$\alpha$}{α}-void cross-correlation}

We used the void catalog presented in our paper~\cite{tomo} to extend galaxy-void correlation studies~\cite{cai} to high redshift $z>2$. To do so, we measured the cross-correlation between the void centers and the Ly$\alpha$~forest pixels used for the tomographic mapping. Our estimator is similar to the one used for the cross-correlation with quasars~\cite{bao}:

\begin{equation}
\xi(A\equiv(r,\mu)) = \frac{\sum\limits_{(i,j)\in A} w_i \delta_{F,i}}{\sum\limits_{(i,j)\in A} w_i},
\end{equation}

\noindent Here $j$ corresponds to void index and $i$ Ly$\alpha$~forest pixel index. The separation between two pairs is characterized by a length $r$ and the angle cosine $\mu$. The weights $w_i$ of the associated $\delta_{F,i}$ pixels depend on noise and redshift (no weights are associated to voids).

\begin{figure}
\begin{minipage}{0.455\linewidth}
\centerline{\includegraphics[trim=6cm 0cm 0cm 0cm,width=0.9\linewidth]{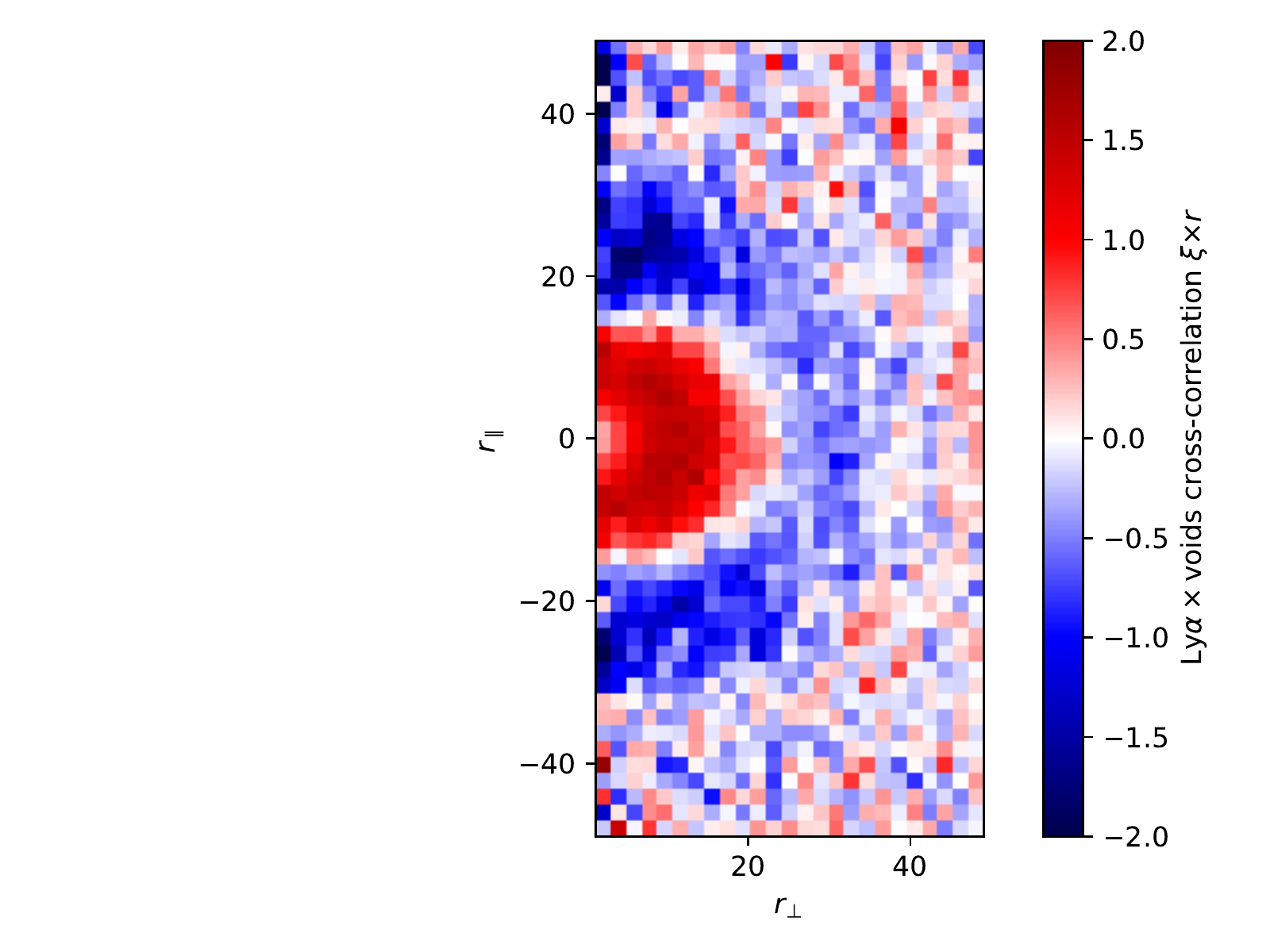}}
\end{minipage}
\hfill
\begin{minipage}{0.535\linewidth}
\centerline{\includegraphics[width=0.9\linewidth]{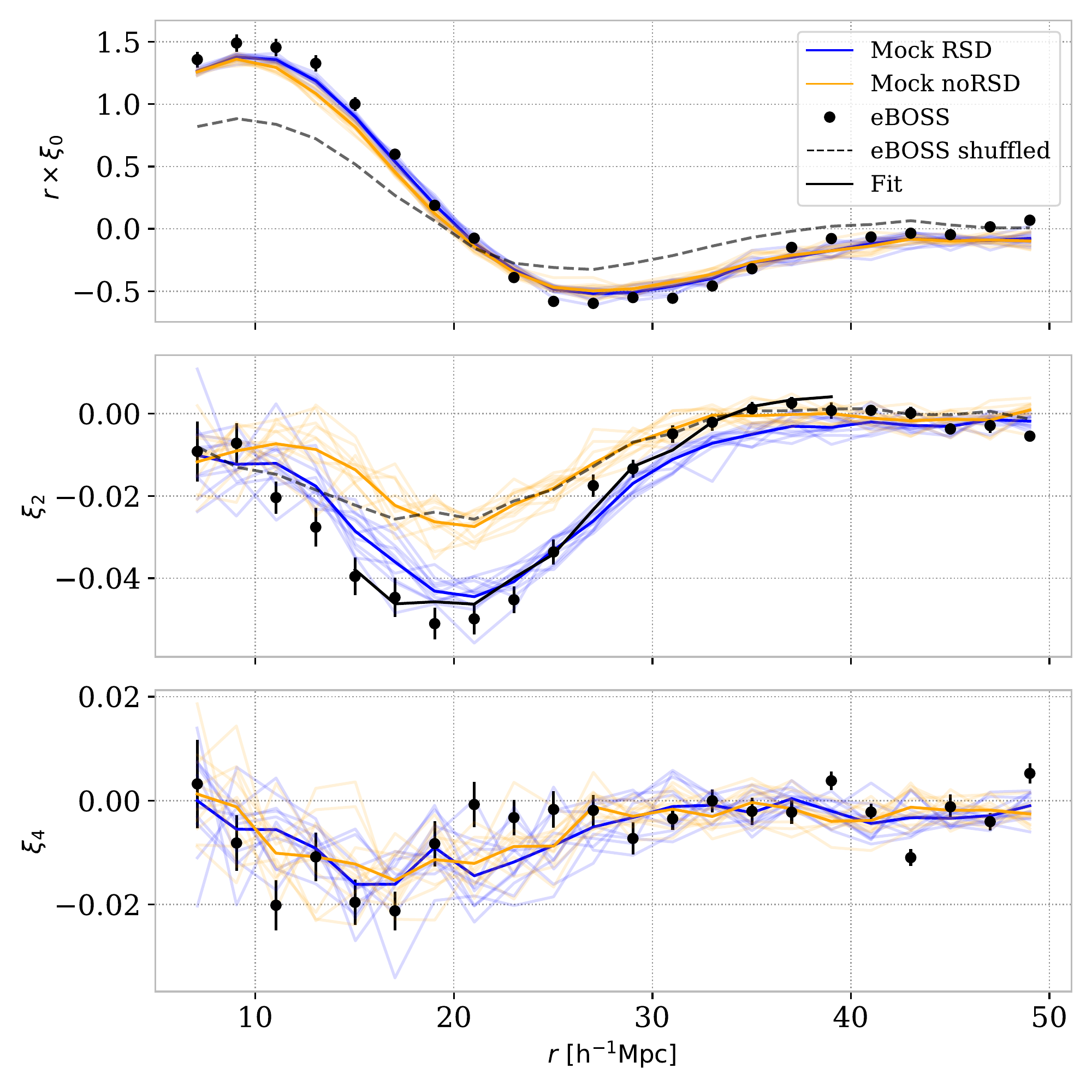}}
\end{minipage}
\caption{(left) Measurement of $r\times \xi(r_{\perp}, r_{\parallel})$ from eBOSS Stripe 82 data. (right) Associated Legendre multipoles for $\ell=0, 2, 4$, for Stripe 82 data (black points) and for mock realizations including RSD (blue) or not (orange). Thin curves represent individual mock realizations, and their average is shown with thick curves. Black dashed curves show the average monopole and quadrupole measured from shuffled data. The black continuous curve shows the fit of the eBOSS quadrupole with Eqn~\ref{eqn:fit_model}.}
\label{fig:xcorr}
\end{figure}

This study aims to use the angular shape of $\xi$ to observe the effect of redshift space distortions (RSD) around voids at redshift $z>2$. The function $\xi$ as computed from Stripe 82 data is illustrated in Fig.~\ref{fig:xcorr} (left0, and a multipole decomposition onto the Legendre polynomial basis, $\xi_{\ell}(r)$ for $\ell=0, 2, 4$, is shown on Fig.~\ref{fig:xcorr} (right). To interpret the measurement, we have also applied this method on a set of mocks similar to the one described above, with the same geometry as Stripe 82. First, we used a series of mocks to evaluate the impact of some instrumental and astrophysical systematics on our measurement (eg. quasar continuum fitting, metals in the IGM). The main conclusion is that for this data set, the impact of the considered systematics is small with respect to statistical fluctuations.

Then, we computed the multipoles of $\xi$ for a set of 10 mocks, as well as for an additional set of companion mocks, labelled "noRSD", for which the effect of velocity flow is not taken into account in the computation of the Ly$\alpha$ absorption. They are shown in Fig~\ref{fig:xcorr} (right). The data clearly demonstrates the existence of the velocity flow, with a statistical significance of $10\,\sigma$.

To interpret this result, we used a simple void model. It is based on the mean velocity profile around voids in the linear regime:

\begin{equation}
    \label{eq:velocity}
    {\bf v} = -\frac{1}{3} \frac{f H}{1+z}\overline{\delta}(r){\bf r}
\end{equation}

This relation links the average velocity ${\bf v}$ to the radial position ${\bf r}$ around void centers, involving the logarithmic growth rate of linear perturbations $f$ and the average isotropic matter density contrast $\overline{\delta}$ inside the sphere of radius $r$. Within this linear model, the monopole and the quadrupole of the Ly$\alpha$-void cross-correlation are connected by a simple relation involving the void RSD parameter $\beta$. This relation is very similar to the case of the galaxy-void correlation~\cite{hamaus}:

\begin{equation}
    \label{eqn:fit_model}
    \xi_2(r) = \frac{2\beta}{3+\beta} \left[ (\xi_0(r) - \overline{\xi_0(r)}) \right]
\end{equation}

As can be seen in Fig~\ref{fig:xcorr} (right), the cross-correlation exhibits a non-zero quadrupole even in the absence of RSD. This feature comes from the particular geometry of the Ly$\alpha$~forest survey, which affects the reconstructed void positions. The average flux contrast of the tomographic map built with the Wiener filter is smaller at locations further away from lines-of-sight. This reduces the efficiency of the void finder, and at the same time, displaces the reconstructed positions of void centers on average, closer towards the nearest line-of-sight with respect to their true positions. We included this effect in a simple way to the model of equation~\ref{eqn:fit_model}. We also shuffled the data to confirm the geometric origin of this effect.

By fitting this corrected model to the Ly$\alpha$-void cross-correlation on data, we obtained an RSD parameter $\beta = 0.52 \pm 0.05$. This value is smaller than that of a similar parameter inferred from the large-scale eBOSS Ly$\alpha$~auto-correlation~\cite{bao}. A full study of the velocity bias from hydrodynamical simulations is probably required to interpret our measurement.

This exploratory work, applied to a statistically limited data set, opens new possibilities for observational cosmology. Upcoming large-field surveys such as WEAVE-QSO~\cite{weave} and DESI~\cite{desi} will extend this measurement to much larger volumes. With an expected line-of-sight density of $\sim 60$~deg$^{-2}$ over a $14,000$~deg$^2$ DESI footprint, the tomographic effect will be reduced, and statistical fluctuations will drastically shrink.

\end{document}